\documentclass[lettersize,journal]{IEEEtran}
\usepackage{amsmath,amsfonts, amssymb}
\usepackage{array}
\usepackage[caption=false,font=normalsize,labelfont=sf,textfont=sf]{subfig}
\usepackage{textcomp}
\usepackage{stfloats}
\usepackage{url}
\usepackage{verbatim}
\usepackage{graphicx}
\usepackage{cite}
\usepackage{threeparttable}
\usepackage{textcomp}
\usepackage{xcolor}
\usepackage{booktabs}
\usepackage{multirow}
\usepackage{siunitx}
\usepackage{stfloats}
\usepackage[ruled,vlined]{algorithm2e}
\begin{document}

\title{Harvesting AI Computation at the Edge via Generic Approximation}

% \author{
% Yihan Wang\textsuperscript{1}, Huiru Yan\textsuperscript{2}, Luxin Zhang\textsuperscript{1}, Long Cheng\textsuperscript{2},\\
% Weiwei Chen\textsuperscript{1}, Ying Wang\textsuperscript{1}, Lei Zhang\textsuperscript{1}, Cheng Liu\textsuperscript{1}, Huawei Li\textsuperscript{1}\\[4pt]
% \textsuperscript{1}\textit{Institute of Computing Technology, Chinese Academy of Sciences}, Beijing, China\\
% \textsuperscript{2}\textit{School of Control and Computer Engineering, North China Electric Power University}, Beijing, China\\
% Email: \{wangyihan24s, luxin1897, chenweiwei46, wangying2009, zlei, liucheng, lihuawei\}@ict.ac.cn, \\
% \{huiruyan, lcheng\}@ncepu.edu.cn
% }

\author{Yihan~Wang,
        Huiru~Yan,
        Luxin~Zhang,
        Long~Cheng,
        Weiwei~Chen, \\
        Ying~Wang,
        Lei~Zhang,
        Cheng~Liu,~\IEEEmembership{Senior Member,~IEEE,}
        Huawei~Li,~\IEEEmembership{Senior Member,~IEEE,}

\thanks{Yihan~Wang, Luxin~Zhang, Weiwei~Chen, Ying~Wang, Lei~Zhang, Cheng~Liu and Huawei~Li are with Institute of Computing Technology (ICT), Chinese Academy of Sciences (CAS), Beijing 100190, China. (e-mail:liucheng@ict.ac.cn)}
\thanks{Huiru~Yan and Long~Cheng are with North China Electric Power University, Beijing 102206, China.}
}



\maketitle

\begin{abstract}

With the widespread adoption of AI in various IoT scenarios such as smart sensing and processing, AI chips have become a common component at the edge. These chips are typically specialized for structured neural network (NN) processing and are designed to meet peak workload demands. However, they are often underutilized and suffer from considerable computational waste due to temporal or spatial redundancy in processing. Conversely, general-purpose processing engines at the edge may struggle with compute-intensive tasks such as signal processing and complex numerical operations because of stringent resource constraints.
To address this imbalance, we propose a framework that harvests unused AI computation resources using general-purpose approximation techniques. The core idea is to automatically convert traditional computing tasks into neural network models via a representative neural architecture search (NAS) method. These approximate versions of general-purpose tasks are then deployed on AI engines during their idle periods. Specifically, we introduce a runtime scheduler that offloads these tasks to AI chips without compromising the performance of primary AI workloads, thereby alleviating the burden on general-purpose processors. Experiments on a representative AIoT processor show that our proposed AI computation harvesting strategy delivers substantial performance improvements across a set of edge processing tasks.

\end{abstract}


\section{Introduction}

With the rapid advancement of AI technologies, IoT devices increasingly rely on specialized AI chips integrated with general purposed processors, often known as AIoT processors. These processors have become the cornerstone of intelligence in modern IoT systems~\cite{atzori2010internet}. However, AI workloads in IoT applications vary dramatically in complexity and resource demands~\cite{jouppi2017datacenter, zeghidour2021leaf, gong2024mcu, wang2025empowering}. For example, neural network (NN) models used for image processing can be orders of magnitude larger than those used for signal processing, such as intelligent sensing~\cite{redmon2017yolo9000, iandola2016squeezenet, howard2017mobilenets}. Even within image processing, models may differ drastically depending on image sizes and task requirements~\cite{tan2019efficientnet}. 

As a result, AI chips, which are typically designed to handle peak workloads, often experience significant under-utilization in the presence of fluctuating computational demands over different NN models and even different layers of a single model, leading to wasted computational resources~\cite{liu2019addressing, raha2024flexnn}. At the same time, general-purpose processors in AIoT systems, constrained by power budgets, often struggle with computation-intensive tasks such as trigonometric functions and data analysis, resulting in substantial processing delays. This mismatch between computation needs and hardware architecture creates inefficiencies in AIoT systems.

Inspired by energy-harvesting techniques that recycle unused energy~\cite{alsharif2024comprehensive}, we propose a novel approach to harvest underutilized AI computation for compute-intensive tasks through NN approximation. This approach replaces traditional arithmetic processing tasks with NN-based approximations and offloads these tasks to AI chips during their idle time. Although NN approximations may require more operations than the original tasks, the massively parallel processing capabilities of AI chips allow them to execute more efficiently than general-purpose processors with much less parallel processing units~\cite{moreau2015snnap, esmaeilzadeh2012neural, eldridge2014neural}. Thus, this strategy not only recycles idle AI computational resources and alleviates the workloads of CPUs, but also potentially achieves higher performance for these tasks.

To enable efficient AI computation harvesting, we adopt the neural architecture search (NAS) to obtain computationally efficient NN models for approximation while adhering to accuracy constraints~\cite{benmeziane2021comprehensive, elsken2019neural, wu2019fbnet, cai2018proxylessnas}. Importantly, we observe that the efficiency of NN approximations is closely tied to the complexity of the task being approximated. Tasks with varying approximation difficulties can lead to inefficiencies if a single NN model is used for the entire task. To address this, we introduce a straightforward yet effective method to automatically decompose computation tasks based on their approximation difficulty. By using larger models for more challenging parts of a task and smaller models for simpler parts, we can reduce the overall overhead of NN approximations with minor accuracy penalty.

In addition to the neural network approximation, we develop a runtime task scheduler that dynamically offloads approximation tasks to the AI engine during its idle periods, ensuring that the AI engine’s primary workload remains unaffected. To validate our approach, we have a set of independent tasks deployed an AIoT processor with the proposed computing harvesting framework. Experimental results demonstrate that the proposed framework achieves significant performance and energy efficiency improvement.

The major contributions of this work can be summarized as follows.
\begin{itemize}
\item We observe the under-utilization of AI computation in AIoT processors and propose a novel AI computation harvesting strategy through generic NN approximation for the first time.

\item We leverage NAS to obtain optimized approximation neural networks for generic computing tasks. In addition, with the observation that the NN approximation efficiency depends on the approximation difficulty, we propose to boost the approximation neural network models by using larger models for more challenging parts of a task and smaller models for simpler parts.

\item On top of the NN approximation, we develop a runtime task scheduler to offload computing-intensive tasks to AI chips and recycle the AI computation without compromising the performance of its primary NN processing.

\item According to our experiments on a set of independent computing tasks, the proposed AI computation harvesting strategy achieves up to 60.5\% higher performance compared to the baseline design framework.

\end{itemize}

\section{Related Work and Motivation}

\subsection{Related Work}
\subsubsection{Edge Computing Optimization}
With the proliferation of AI workloads at the edge, deep neural networks, while delivering high prediction accuracy, face stringent energy and resource constraints in embedded deployments. Consequently, extensive research has targeted software‑level optimizations for edge processors. Representative efforts include the design of bespoke neural network operators that align with edge processor architectures~\cite{lin2021memory}, low–bit‑width quantization and structured pruning to curtail compute and memory demands~\cite{kuzmin2023pruning, le2023efficient}, and hybrid cloud–edge co‑execution schemes for balanced workload distribution~\cite{xu2023survey}. Despite these advancements, software‑only approaches exhibit inherent limitations, driving the widespread adoption of specialized Deep Learning Accelerator (DLA) to meet stringent performance and efficiency requirements~\cite{alam2024survey}.
Nonetheless, DLA utilization remains suboptimal due to the highly variable compute demands across different models, individual layers, and temporal phases of execution. This variability results in significant periods of idle hardware and wasted compute capacity~\cite{sze2017efficient, zhang2023shepherd, kwon2018maeri}. Meanwhile, general‑purpose processors, although capable of diverse tasks such as signal processing and data analytics, are inefficient for the multiply-accumulate-dominated workloads typical of neural networks~\cite{chen2016eyeriss}. Moreover, DLA, built around arrays of MAC units and optimized for regular operations (e.g., matrix multiplications and convolutions), cannot be readily repurposed for non‑AI tasks\cite{kao2022formalism, jouppi2023tpu}. This persistent mismatch between computational demand profiles and available hardware resources is generally overlooked, leaving substantial headroom for improved resource harvesting and utilization.

\subsubsection{Generic Computing Approximation}
Neural networks have been established as universal function approximators since the seminal work of Cybenko and Hornik in the late 1980s and early 1990s, which showed that multilayer feedforward architectures can approximate any continuous mapping given sufficient width or depth~\cite{cybenko1989approximation,hornik1991approximation}. While these theoretical guarantees require idealized, often impractically large networks, more recent research has quantified how depth and width trade off against approximation error under realistic resource constraints. For example, Yarotsky proved that deep ReLU networks need only logarithmic depth in the target error to retain tight approximation bounds~\cite{yarotsky2017error, siegel2023optimal}, and other studies have demonstrated that depth‑efficient architectures can reduce parameter counts exponentially compared to shallow models~\cite{lu2017expressive, vardi2020neural, petersen2018optimal}.
Although deploying generic approximation networks on general‑purpose processors remains costly, the real advantage emerges when these networks run on specialized DLA optimized for the regular multiply‑accumulate patterns inherent in neural models. By recasting arbitrary functions as neural approximators, one effectively transforms irregular compute into a stream of highly optimized matrix and convolution operations, unlocking the high throughput of modern AI engines even for non‑AI workloads. To tailor these approximators to a given accelerator’s characteristics, lightweight architecture search methods—such as gradient‑based DARTS variants~\cite{liu2018darts, liang2019darts+, zela2019understanding, chu2020fair}—can be used to discover compact, high‑accuracy networks without prohibitive search cost. This generic approximation methodology thus provides a principled mechanism for closing the gap between diverse edge compute demands and the fixed resources of DLA.

\subsection{Motivation}
Edge applications typically demand a diverse set of neural networks because no single model can address every task, so resource utilization on AIoT processors varies widely across the model architectures. To quantify this, we deployed four representative networks (see Table~\ref{tab:Model utilization deployed on MAX78000}) on a typical AIoT processor MAX78000~\cite{moss2022ultra} from Microchip and calculated the utilization with Equation~\ref{eq:utilization}. Note that $Lat$ represents average network inference latency, $T_{OPS}$ represents the total number of operations required by each neural network, and $P_{OPS}$ represents the total number of operations executed at peak performance.

\begin{equation}
U = \frac{\mathrm{Lat} \times \mathrm{T}_{\mathrm{OPS}}}{\mathrm{P}_{\mathrm{OPS}}}
\label{eq:utilization}
\end{equation}

Our measurements show that, except for Tiny‑YOLOv2, the MAX78000’s compute utilization is generally below 50 \%. Furthermore, most IoT workloads require only 10–15 fps, introducing additional temporal slack. When both spatial underutilization (low average utilization) and temporal slack (lower frame‑rate requirements) are considered, a substantial fraction of the AIoT processor’s capacity remains idle, leaving a large design space for harvesting this unused computation.

\begin{table}[htbp]
\caption{Computing Resource Utilization of A Typical AIoT Processor (MAX78000)}
    \begin{center}
        \begin{tabular}{cccc}
        \toprule
        Model & GOPS(Infer) & GOPS(30FPS) & Utilization \\
        \midrule
        Tiny-YOLOv2 & 2.08 & 62.4 & 96.0\% \\
        SqueezeNet & 0.8 & 24 & 36.9\% \\
        MobileNet & 1.2 & 36 & 55.4\% \\
        EfficientNet-Lite0 & 0.6 & 18 & 27.7\% \\
        \bottomrule        \end{tabular}
    \label{tab:Model utilization deployed on MAX78000}
    \end{center}
\end{table}
\section{AI Harvesting Framework}
\begin{figure*}[t]
\centerline{\includegraphics[scale=0.9]{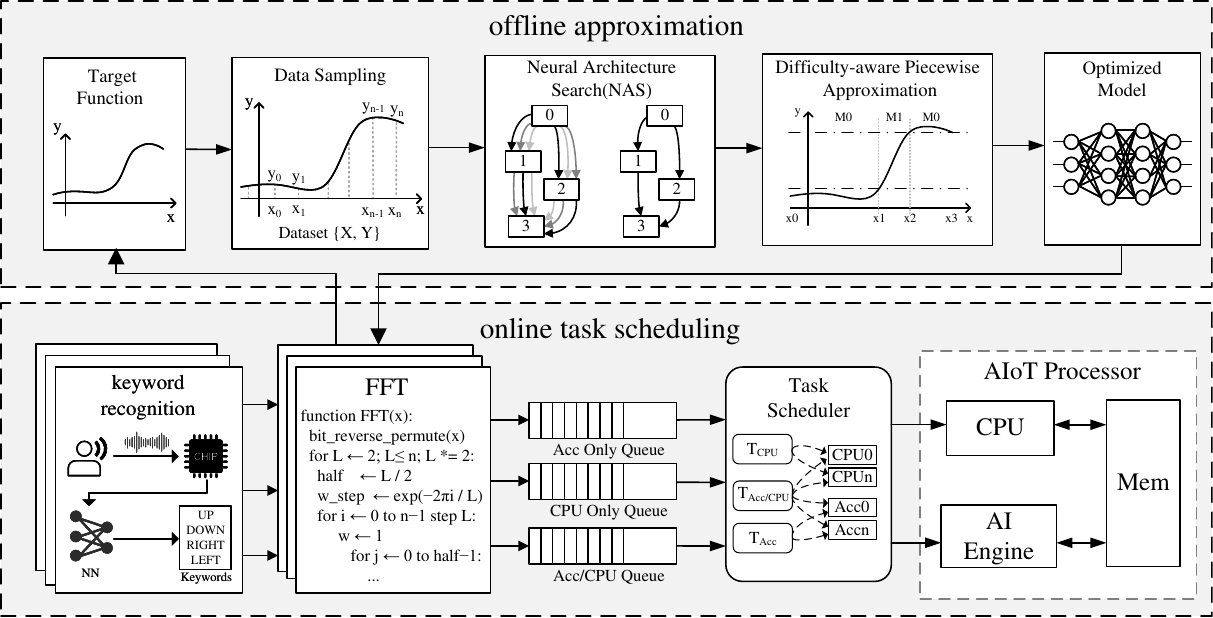}}
\caption{AI Harvesting Framework.}
\label{fig:AI_Harvesting_Framwork}
\end{figure*}
In this work, we propose an end‑to‑end AI computation harvesting framework that systematically transforms non‑AI workloads into optimized neural approximators and dynamically allocates all tasks across an AIoT processor with a closely coupled general‑purpose processor and a DLA. The framework as shown in Fig.~\ref{fig:AI_Harvesting_Framwork} consists of an offline approximation part and an online task scheduling part. 
% In an offline part, each target function whether signal processing, control logic, or data analysis, is approximated by a compact neural network discovered via lightweight Neural Architecture Search (NAS). By tailoring depth, width, and operator sets to the characteristics of the DLA, these approximators retain accuracy while conforming to strict resource and latency budgets. 
In an offline part, each target function whether signal processing, control logic, or data analysis, is approximated by a compact neural network discovered via lightweight NAS. By tailoring depth, width, and operator sets to the characteristics of the DLA, these approximators retain accuracy while conforming to strict resource and latency budgets. 
At runtime, a heuristic scheduler orchestrates both conventional code and the pre‑built neural network models, expanding the scheduling design space far beyond traditional CPU‑only or DLA‑only pipelines. Thanks to the generic neural approximators, the scheduler can flexibly assign compute‑intensive segments to the AI engine where matrix‑centric operations enjoy peak throughput and offload control‑oriented or irregular tasks to the CPU. This unified approach not only maximizes end‑to‑end resource utilization but also effectively bridges the gap between dynamic workload demands and fixed hardware capabilities, delivering significant gains in overall system efficiency.

\subsection{NAS-based Neural Network Approximation}
To enable the efficient deployment of non-AI computing tasks on DLA, we propose approximating target functions with compact neural network models. Traditional hand-crafted designs for such approximators are error-prone and often require substantial AI expertise. Moreover, naive use of neural networks typically introduces significant computational overhead, making them impractical for lightweight or latency-sensitive applications. To overcome these limitations, we leverage NAS to automate the design of efficient, accurate, and deployable neural approximators.
Specifically, we adopt the DARTS framework to generate neural network models that meet user-defined constraints on accuracy, floating-point operations (FLOPs), and model size. It aims to search for a neural network model $\lambda$ that minimizes the mean squared error (MSE) function $f_{MSE}$ over the search space $\chi$, subject to hardware-aware constraints $C = \{C_{flops}, C_{size}\}$:

\begin{equation}
    \lambda^* = \mathop{\arg\min}_{ \lambda \in \chi} { f_{MSE}(C) }
\end{equation}

The DARTS-based NAS consists of two stages: (1) constructing a supernet from candidate cell architectures in the search space and training it to select the best-performing cell; (2) reconstructing the final architecture from the derived genotype and training the complete model. The final model $M$ is parameterized by $M:\{C, L, S\}$, where $C$ is the number of channels per cell, $L$ is the number of stacked layers and $S$ is the number of nodes per cell ($S \geq 2$). 

To improve search stability and final model quality, we apply the Fair-DARTS optimization technique to address two major issues in the original DARTS: (i) architectural collapse due to overuse of skip connections, and (ii) mismatch between continuous and discrete representations during search. We apply a sigmoid activation $\sigma$ to each continuous variable $\alpha_{i;j}$ and introduce an early stopping mechanism to stabilize training. The loss function use FairDARTS method to helps reduce discrepancies when discretizing the architecture. We also tailor the DARTS operator set to better suit approximation tasks. While DARTS originally employs convolutional operations optimized for classification, we find that linear layers yield better performance in approximating mathematical functions. Therefore, we redefine the base operations to consist of linear layers followed by activation functions. This design not only improves approximation accuracy but also aligns well with typical DLA pipelines, which are often optimized for matrix-vector computations involving linear and activation layers.

We notice that the difficulty of neural network approximation varies across different segments of a function. Smooth regions can be effectively approximated using small, shallow models, while more complex regions require deeper networks for sufficient accuracy. Relying on a single model for the entire function often leads to suboptimal trade-offs between computational efficiency and accuracy. To address this, we propose a difficulty-aware piecewise approximation as shown in Algorithm \ref{alg:partition_intervals_vec}. It partitions the input space based on approximation error and assigns specialized neural networks to each sub-interval. Lightweight models are used in regions where simple approximations suffice, while more complex sub-models are employed in challenging areas. The models are then combined into a hybrid architecture that dynamically adapts to input complexity. This approach enables general-purpose function approximation on DLA with flexible trade-offs between accuracy and efficiency, thereby maximizing hardware utilization.

\begin{algorithm}[htbp]
  \caption{difficulty-aware piecewise approximation}
  \label{alg:partition_intervals_vec}
  \KwIn{Absolute-error array $E[0 \textbf{ to } N-1]$, threshold list $T$}
  \KwOut{List of intervals $(\mathit{start},\mathit{end},\mathit{label})$}

  Sort $T$ in ascending order\;
  $labels \gets \text{digitize}(E, T)$\;
  $N \gets |labels|$\;
  \If{$N = 0$}{
    \Return $\emptyset$\;
  }
  % \tcp{Add sentinel elements at both ends}
  $padded \gets [\,labels[0]\,] + labels + [\,labels[N-1]\,]$\;
  % \tcp{Compute adjacent differences}
  $\Delta \gets [\,padded[i+1] - padded[i]\;\mid\; i=0 \textbf{ to } N\,]$\;
  $change\_indices \gets \{\,i \mid \Delta[i] \neq 0\}$\;
  $starts \gets change\_indices[$0$ \textbf{ to } |change\_indices|-2]$\;
  $ends   \gets change\_indices[1 \textbf{ to } |change\_indices|-1]$\;

  $intervals \gets []$\;
  \For{$k \gets 1$ \KwTo $|starts|$}{
    $s \gets starts[k]$\;
    $e \gets ends[k]$\;
    append $(s, e, labels[s])$ to $intervals$\;
  }
  \Return $intervals$\;
\end{algorithm}

In addition, we also have some minor optimization applied to the NAS. To support a wide range of function approximation needs, we sample input-output pairs from the target computing functions to build training datasets for NAS. Typically, an evenly distributed sampling is sufficient, but a gradient-aware non-evenly distributed sampling will be beneficial when the target function fluctuates abruptly. While DLA exhibit varied computing efficiency over different neural network operators, we incorporate a hardware-awareness component into the NAS framework, ensuring that the selected architectures are optimized for the specific execution characteristics of the target DLA. 

Finally, inspired by general-purpose computing libraries such as ARM's CMSIS-DSP, we build a generalized neural approximation library for DLA. This library provides a suite of pre-trained neural network models that approximate commonly used computing functions. Supported by compilation and deployment toolchains tailored for different DLA architectures, this library enables direct, high-performance execution of traditional workloads on AI hardware.

\subsection{Runtime Task Scheduling on AIoT Processors}
This section mainly investigates the task scheduling algorithm on a CPU-DLA architecture, which can offload intensive yet irregular computing tasks to DLA based NN approximation in previous section such that the wasted computing capability of DLA can be recycled. The scheduler adopts a heuristic strategy that considers both task latency and resource availability. Tasks are prioritized to execution units (CPU or DLA) based on which can complete them the earliest. 
For tasks offloaded to the DLA, the model loading overhead i.e. the data transfer latency from CPU to DLA is also taken into account. The task scheduling can be formulated as follows.

Let the task set be $\left \{T_0, T_1, ... , T_{M-1}\right \}$, and the processor set be $\left \{0, 1, ... , {D-1}\right \}$, where $0$ denotes the CPU, and $\left \{1:D-1\right \}$ are DLA. The type of tasks $\tau_i \in \left \{T^{CPU-only}, T^{DLA-only}, T^{DLA/CPU} \right \}$.
Each task can be scheduled on exactly one processor:
\begin{equation}
\sum_{j=0}^{D-1} a_{i,j} =1  \ \  \forall i\in [0,M)
\end{equation}
where $a_{i,j}=1$ represents $i-th$ task is sent to $j-th$ processor (CPU or DLA) for execution, $M$ and $D$ denote the number of tasks and processors. Each task has an execution time on different processors, and for the tasks assigned to the DLA, the time of loading model data from the CPU to the DLA also needs to be considered.
\begin{equation}
\begin{split}
    E_{i,j}= 
    \left\{\begin{matrix}
  E^{C}_i & j=0 \\
  E^{DLA}_i & else
\end{matrix}\right.
\end{split}
\end{equation}
\begin{equation}
\begin{split}
    L_{i,j}= 
    \left\{\begin{matrix}
  0 & j=0 \\
  L_{i,j} & else
\end{matrix}\right.
\end{split}
\end{equation}
where $E_{i,j}$ denotes the executing time of the task $i$ on the processor $j$, which $E^{C}_i$ and $E^{DLA}_i$ represent the processing time of task $i$ on the CPU and DLA, respectively. $L_{i,j}$ represents the loading time of the task $i$ to the processor $j$, and $j=0$ denotes that the processor is CPU. During the process of loading tasks into the DLA, a transmission serialization constraint exists, meaning that the system permits only one task to perform transmission operations at any given moment. Thus, we set a global transfer time $Trans$, which represents the start transfer time of the tasks assigned to the DLA. We define the scheduling triplet of task $T_i$ as:
\begin{equation}
S_i = (a_{i,j}, s_{i}, f_{i})
\end{equation}
where $a_{i,j}$, $s_{i}$, $f_{i}$ denote the processor to which the task is assigned, the start execution time and the completion time of the task $i$, respectively.

In the scheduling process, we adopt the greedy scheduling strategy based on earliest finish time (EFT), as shown in Algorithm \ref{alg:EFT}. Specifically, for the task whose type $\tau_i$ is $T^{CPU-only}$, it is only assigned to the CPU (set $j=0$), for the other two types of tasks to be scheduled, it is calculated separately on all its optional processors (for the task whose $\tau_i$ is $T^{DLA-only}$, j $\in$ [1,D-1)):
\begin{equation}
s_i = \begin{cases} \max (EST_i, D_j) & j=0 \\ 
      \max (Trans, D_j) & else \end{cases}
\end{equation}
\begin{equation}
f_i = s_i + E_{i,j}
\end{equation}
\begin{equation}
EST_i = \max_{T_k \in Dep(i)} f_k
\end{equation}
where $EST_{i}$ and $Dep(i)$ denote the earliest start time and the  set of predecessor tasksof the task $i$, respectively. Then, according to the task completion time $f_{i}$, EFT algorithm gives priority to assigning the task to the execution processor that can complete the task earliest, and updates the the ready time of processor $j$ and global transfer time. 
\begin{equation}
D_j = f_i \ \ \ if \ a_{i,j}=1
\end{equation}
\begin{equation}
Trans = \max (EST_i, Trans) +L_{i,j}
\end{equation}
The maximum value of completion time among all processors determines the total runtime of this batch of tasks. Therefore, the optimization goal is to minimize the earliest completion time of processors through scheduling.
\begin{equation}
    min(\max_j(D_{j})) 
\end{equation}

\begin{algorithm}[htbp]
\caption{scheduling based on earliest finish time (EFT)}
\label{alg:EFT}
\KwIn{Task set $\{T_0, T_1, \dots, T_{M-1}\}$, processor set $\{0, 1, \dots, D-1\}$ (0 is CPU, others are DLAs), Execution time $E_{i,j}$, Loading time $L_{i,j}$, Task type $\tau_i$}
\KwOut{Task schedule $S_i = (a_{i,j}, s_i, f_i)$ for each $T_i$}

Initialize ready time of processor $D_j \gets 0 \ \forall j \in [0,D)$\;
Initialize global transfer start time $Trans \gets 0$\;

\For{$i \gets 0$ \KwTo $M-1$}{
    $EST_i \gets \max\limits_{T_k \in Dep(T_i)} f_k$  \;
    \eIf{$\tau_i = T^{CPU-only}$}{
        $a_{i,0} \gets 1$; $s_{i} \gets \max(EST_i, D_0)$; $f_{i} \gets s_{i} + E_{i,0}$\;
    }{
        Initialize assigned processor $p \gets \text{None}$; earliest finish time $eft \gets +\infty$ \;
        \For{each processor $j \in D$}{
            \eIf{$j = 0$}{
            $start\gets \max(EST_i, D_0)$\; $finish\gets start + E_{i,0}$;
            }{
            $trans \gets \max (EST_i, Trans) + L_{i,j}$\;
            $start\gets \max(trans, D_j)$\;
            $finish\gets start + E_{i,j}$\;
            }
            \If{$finish < eft$}{
            $p \gets j$\;
            $eft \gets finish$;
            }
        }
        Assign $T_i$ to processor $p$: $a_{i,p} \gets 1$ \;
        $s_{i} \gets start$; $f_{i} \gets eft$; $D_{p} \gets f_{i}$ \; 
        Update $Trans \gets trans \ \ \ if \ p > 0$ \; 
    }
}
Optimization objective: $\min \mathrm{Makespan} = \max_j D_j$\;
\Return{$\{S_i = (a_{i,j}, s_{i}, f_{i})\}_{i=0}^{M-1}$, $Makespan$}
\end{algorithm}

\section{Experiments}

\subsection{Experiment Setups}

\subsubsection{Computing Platform Setups}
In this work, we use the MAX78000 AIoT processor from Analog Devices (ADI) as a representative platform to demonstrate the effectiveness of the proposed AI harvesting framework. The MAX78000 features an Arm Cortex‑M4F core running at 100 MHz and an on-chip DLA operating at 60 MHz. The DLA consists of four groups of parallel units sharing common control logic. Hence, it can be considered as four accelerator units and each of them can be shutdown independently to fit the application requirements. 

\subsubsection{Kernel Benchmark}
To assess the generality of computing approximations at the edge, we construct a benchmark suite that includes core mathematical functions from the C standard library, representative signal processing algorithms, and several lightweight neural network models, as summarized in Table~\ref{tab:benchmark}. These benchmarks are widely used in practice to support a broad range of high-level applications and require both computational efficiency and accuracy. 

\begin{table*}[htbp]
\caption{Kernel Benchmark}
    \begin{center}
    \resizebox{0.8\linewidth}{!}{
        \begin{tabular}{cccccccccc}
        \toprule
        \textbf{Task} & \textbf{Input Range} & \textbf{MSE (FP32)} & \textbf{MSE(int8)} & \textbf{CPU(us)} & \textbf{DLA(us)} & \textbf{\# of OPs} & \textbf{Loading(us)}  \\
        \midrule
        32 FFT     & $(-100, +100)$  & \num{2.90e-12} & \num{2.76e-3} & 2588  & 0.256 & 49408  & 35.6  \\
        64 FFT     & $(-100, +100)$  & \num{3.42e-10} & \num{3.47e-2} & 6668  & 0.612 & 148096 & 105   \\
        128 FIR     & $(-100, +100)$ & \num{4.06e-2}  & \num{1.05e-1} & 47800 & 0.504 & 114944 & 82.0   \\
        128 Biquad  & $(-100, +100)$ & \num{1.14e-4}  & \num{5.24e-3} & 858   & 0.503 & 114944 & 82.0   \\
        sin        & $(-2\pi, 2\pi)$      & \num{1.27e-6}  & \num{4.31e-4} & 24.9 & 0.055 & 205824 & 4.47 \\
        cos        & $(-2\pi, 2\pi)$      & \num{3.69e-5}  & \num{4.00e-4} & 25.0 & 0.041 & 137216 & 3.37 \\
        tan        & $(-\pi/4, \pi/4)$    & \num{2.33e-6}  & \num{2.27e-4} & 43.0 & 0.054 & 204800 & 4.86 \\
        asin       & $(-1,1)$             & \num{2.54e-6}  & \num{1.03e-4} & 3.46 & 0.027 & 69632  & 1.89 \\ 	
        acos       & $(-1,1)$             & \num{1.85e-6}  & \num{1.10e-4} & 3.36 & 0.054 & 200704 & 4.86 \\
        atan       & $(-1,1)$             & \num{5.18e-7}  & \num{2.20e-4} & 28.5 & 0.027 & 69632  & 1.89 \\
        sinh       & $(-1,1)$             & \num{3.81e-3}  & \num{2.26e-3} & 4.92 & 0.033 & 101376 & 2.38 \\
        cosh       & $(-1,1)$             & \num{2.15e-5}  & \num{7.45e-3} & 5.36 & 0.054 & 200704 & 4.86 \\
        tanh       & $(-10,10)$           & \num{1.25e-7}  & \num{8.07e-5} & 0.93 & 0.048 & 200704 & 4.12 \\
        asinh      & $(-1,1)$             & \num{1.23e-7}  & \num{1.96e-4} & 4.59 & 0.041 & 135168 & 3.38 \\
        acosh      & $(1,3)$              & \num{1.33e-5}  & \num{7.22e-4} & 47.7 & 0.027 & 69632  & 1.89 \\
        atanh      & $(1,3)$              & \num{1.25e-5}  & \num{1.68e-4} & 4.54 & 0.054 & 167936 & 4.87 \\
        exp	       & $(-1,1)$             & \num{9.54e-6}  & \num{1.61e-2} & 5.87 & 0.046 & 169984 & 4.12 \\
        ln	       & $(0.1,5)$            & \num{2.94e-7}  & \num{3.80e-4} & 28.1 & 0.046 & 200704 & 4.86 \\
        minist net & -                    & -              & - & -         & 1409 & 10883968 & 4882.81  \\
        kw20 net   & -                    & -              & - & -         & 1844 & 8402528  & 12084.96 \\
        cifar-10   & -                    & -              & - & -         & 4570 & 36481536 & 20874.02 \\
        cifar-100  & -                    & -              & - & -         & 4584 & 36527616 & 24072.26 \\
        \bottomrule
        \end{tabular}}
    \label{tab:benchmark}
    \end{center}
\end{table*}

\subsubsection{Application Benchmark} 
To comprehensively evaluate the proposed approximation approach, we employ both realistic and synthetic applications as benchmarks. The realistic applications consist of a motor control algorithm (Field Oriented Control, FOC) and a gyro-magnetometer fusion algorithm (EKF-based heading estimation, Heading). The synthetic applications are constructed from random task graph models, enabling coverage of more complex and diverse application scenarios.

\textbf{Realistic Applications:} 
The first realistic application is Field Oriented Control (FOC), a mainstream control strategy for permanent magnet synchronous motors (PMSMs) and induction motors. FOC decouples the regulation of direct- and quadrature-axis currents through Clarke and Park transformations, followed by inverse Park transformation and space vector pulse width modulation (SVPWM) to generate three-phase driving signals. In our experiments, two neural network models were trained: FOC-NN1, which approximates the Clarke and Park transformations
(Eqs.\eqref{eq:clarke}, \eqref{eq:park}), and FOC-NN2, which approximates the inverse Park and SVPWM operations (Eqs.\eqref{eq:invpark}, \eqref{eq:svpwm}). 

% clarke
\begin{equation}
\begin{bmatrix}
i_\alpha \\
i_\beta
\end{bmatrix}
=
\frac{2}{3}
\begin{bmatrix}
1 & -\tfrac{1}{2} & -\tfrac{1}{2} \\
0 & \tfrac{\sqrt{3}}{2} & -\tfrac{\sqrt{3}}{2}
\end{bmatrix}
\begin{bmatrix}
i_a \\
i_b \\
i_c
\end{bmatrix}
\label{eq:clarke}
\end{equation}

% park
\begin{equation}
\begin{bmatrix}
i_d \\
i_q
\end{bmatrix}
=
\begin{bmatrix}
\cos\theta_e & \sin\theta_e \\
-\sin\theta_e & \cos\theta_e
\end{bmatrix}
\begin{bmatrix}
i_\alpha \\
i_\beta
\end{bmatrix}
\label{eq:park}
\end{equation}

% inverse Park
\begin{equation}
\begin{bmatrix}
v_\alpha \\
v_\beta
\end{bmatrix}
=
\begin{bmatrix}
\cos\theta_e & -\sin\theta_e \\
\sin\theta_e & \cos\theta_e
\end{bmatrix}
\begin{bmatrix}
v_d \\
v_q
\end{bmatrix}
\label{eq:invpark}
\end{equation}

% SVPWM
\begin{equation}
\left\{
\begin{aligned}
v_a &= v_\alpha \\
v_b &= -\tfrac{1}{2}v_\alpha + \tfrac{\sqrt{3}}{2} v_\beta \\
v_c &= -\tfrac{1}{2}v_\alpha - \tfrac{\sqrt{3}}{2} v_\beta
\end{aligned}
\right.
\label{eq:svpwm}
\end{equation}

The second application is extended Kalman filter (EKF)-based heading estimation, which fuses gyroscope and magnetometer data to provide robust orientation tracking. This algorithm is commonly used in embedded and robotic systems to mitigate sensor noise and drift, thereby ensuring reliable heading estimation in real time.

\begin{equation}
\left\{
\begin{aligned}
m_x^c &= m_x \cos\theta + m_y \sin\phi \sin\theta + m_z \cos\phi \sin\theta \\
m_y^c &= -m_y \cos\phi + m_z \sin\phi \\
\psi_{\text{mag}} &= \arctan2(-m_y^c,\, m_x^c)
\end{aligned}
\right.
\label{eq:tilt_heading}
\end{equation}

\textbf{Synthetic Applications:}
To further evaluate the effectiveness of the AI harvesting strategy in complex scenarios involving multiple interdependent tasks, we synthesized four representative applications-Syn-App1, Syn-App2, Syn-App3, and Syn-App4—each composed of 50 computing tasks randomly selected from Table~\ref{tab:benchmark}. 
Each synthetic application is constructed as a directed acyclic graph where tasks are randomly sampled and the task dependencies are randomly generated following a topological order ($T_i$ depends only on $T_j$ where j \textless  i). For each task, up to 2–3 parent nodes are randomly selected with a connection probability that is iteratively adjusted to maintain an average node degree of 1.0 ± 0.2. Cycles, if generated, are automatically removed to avoid cyclicity.
To better emulate realistic workloads, only a subset of tasks in each application is designed to be approximal with neural networks and can therefore be executed on both CPUs and DLAs, while the remaining tasks are restricted to CPU execution. Specifically, Syn-App1 contains 10\% approximal tasks, Syn-App2 contains 30\%, Syn-App3 contains 50\%, and Syn-App4 contains 80\%.  
% We assume that computing tasks arrive in a streaming fashion, and the scheduler dynamically determines their execution order to optimize overall performance. 
% Each synthetic workflow is constructed as a directed acyclic graph where tasks are randomly sampled from 18 predefined templates and classified based on the approximability ratio. 
% Task dependencies follow topological ordering (T\_i depends only on T\_j where j < i). We control the task graph density by maintaining an average node degree of 1.2 ± 0.2. The maximum number of dependencies per task is limited to 2-3. 
% Each workflow is represented as a directed acyclic graph (DAG), where task dependencies are randomly generated following a topological order (T\_i depends only on T\_j where j < i). For each task, up to 2–3 parent nodes are randomly selected with a connection probability that is iteratively adjusted to maintain an average node degree of 1.0 ± 0.2. Cycles, if generated, are automatically removed to ensure acyclicity.

\subsubsection{Approximation Setups}
In this experiment, we have the computing tasks approximated with neural network models via NAS. The approximate neural network models are trained using stochastic gradient descent (SGD) with momentum 0.9, an initial learning rate of 0.025 (annealed via a cosine schedule over 50 epochs down to 0.001), and a batch size of 32. The neural architecture search phase spans 50 search epochs, followed by a reconstruction phase of up to 400 epochs with early stopping (patience set to 15 epochs). We use mean squared error (MSE) as the primary accuracy metric.

\subsection{Kernel Approximation Evaluation}
\subsubsection{Performance Evaluation}
The runtime performance of the benchmark suite on both the CPU and DLA is summarized in Table~\ref{tab:benchmark}. In general, the model size primarily depends on the parameters of the search model and the configuration of the search space, which mainly includes cascaded linear and activation layers, skip connections, and “None” operations. For simple approximation tasks, the search process tends to favor shallow models composed of linear and activation layers to minimize the mean squared error (MSE). In such cases, the DLA is severely underutilized, resulting in negligible differences in parameter counts among the approximation models. Notably, the execution time on the AI accelerator remains almost identical even when the models differ slightly, further confirming this observation. However, we also notice that model loading from flash to RAM introduces a non-trivial overhead compared to the actual accelerator execution time, becoming a potential performance bottleneck.

Although the overall performance gains from neural approximation over CPU execution are modest for simple tasks, the results still underscore the potential of neural network–based approximation. Kernel execution on the AI accelerator consistently outperforms its CPU counterpart, and the model loading overhead can be effectively amortized through data-level parallelism. Moreover, more complex computational workloads that aggregate multiple kernel functions are expected to benefit more significantly, offering greater opportunities for performance enhancement through approximation. The performance speedups reported for signal processing tasks in Table~\ref{tab:benchmark} further corroborate this trend.

\subsubsection{Approximation Accuracy Evaluation}
As shown in Table~\ref{tab:benchmark}, the proposed approximation method generally achieves low mean square error (MSE), particularly for transition functions. However, for a few functions with highly variable input–output relationships, the neural network approximation exhibits relatively higher MSE. As expected, quantization further increases the error, yet the overall MSE typically remains around $10^{-3}$, which is acceptable for most embedded applications. For kernel functions exhibiting substantial variation, we employ a difficulty-aware piecewise model to enhance approximation accuracy. This piecewise model partitions the target function into multiple segments based on their approximation difficulty, enabling trade-offs between accuracy and computational overhead. In this experiment, five representative mathematical functions listed in Table~\ref{tab:Compare sub-model in similar acc} are approximated using three models: a small model ($Model_S$), a large model ($Model_L$), and a piecewise model ($Model_P$). The results show that $Model_P$ achieves accuracy comparable to that of the largest model ($Model_L$) while significantly reducing the number of operations. However, this approach introduces additional memory overhead, as multiple sub-models must be stored. Interestingly, in some cases—such as the $tan$ function—the piecewise model even surpasses the larger model in both accuracy and efficiency. This observation highlights that increasing model size does not necessarily guarantee higher approximation accuracy, a phenomenon further demonstrated in the subsequent NAS-based approximation analysis.

\begin{table}[htbp]
\caption{Compare sub-model in similar acc}
\centering
\resizebox{1\linewidth}{!}{
\begin{tabular}{cccccc}
\toprule
Fun & Input Range & Model & MSE & Flops & Model Size(K) \\
\midrule
\multirow{3}{*}{sin} & \multirow{3}{*}{$(-2\pi, 2\pi)$} & $Model_S$ & \num{7.68e-5} & 166912 & 5.409 \\
                     &                            & $Model_L$ & \num{2.34e-7} & 691200 & 22.273 \\
                     &                            & $Model_P$ & \num{2.67e-7} & 570400 & 37.315 \\
\midrule
\multirow{3}{*}{cos} & \multirow{3}{*}{$(-2\pi, 2\pi)$} & $Model_S$ & \num{3.69e-5} & 156910 & 6.465 \\
                     &                            & $Model_L$ & \num{8.90e-6} & 691200 & 22.273 \\
                     &                            & $Model_P$ & \num{8.91e-6} & 516684 & 39.427 \\
\midrule
\multirow{3}{*}{tan} & \multirow{3}{*}{$(-\pi/4, \pi/4)$} & $Model_S$ & \num{2.33e-6} & 156910 & 7.465  \\
                     &                            & $Model_L$ & \num{4.59e-7} & 691200 & 22.273 \\
                     &                            & $Model_P$ & \num{4.99e-7} & 330752 & 10.689 \\
\midrule
\multirow{3}{*}{atan} & \multirow{3}{*}{$(1,3)$} & $Model_S$ & \num{5.18e-7} & 199680 & 6.465  \\
                      &                            & $Model_L$ & \num{3.97e-7} & 691200 & 25.632 \\
                      &                            & $Model_P$ & \num{4.01e-8} & 623625 & 42.786 \\
\midrule
\multirow{3}{*}{ln}  & \multirow{3}{*}{$(0.1,5)$} & $Model_S$ & \num{1.35e-7} & 199680 & 6.465  \\
                      &                            & $Model_L$ & \num{5.69e-8} & 691200 & 22.273 \\
                      &                            & $Model_P$ & \num{5.57e-8} & 686610 & 39.423 \\
                      
\bottomrule
\end{tabular}
}
\label{tab:Compare sub-model in similar acc}
\end{table}

\begin{figure*}[htbp]
\centering
\subfloat[Channel Analysis.\label{fig:channel}]{
  \includegraphics[width=0.31\linewidth]{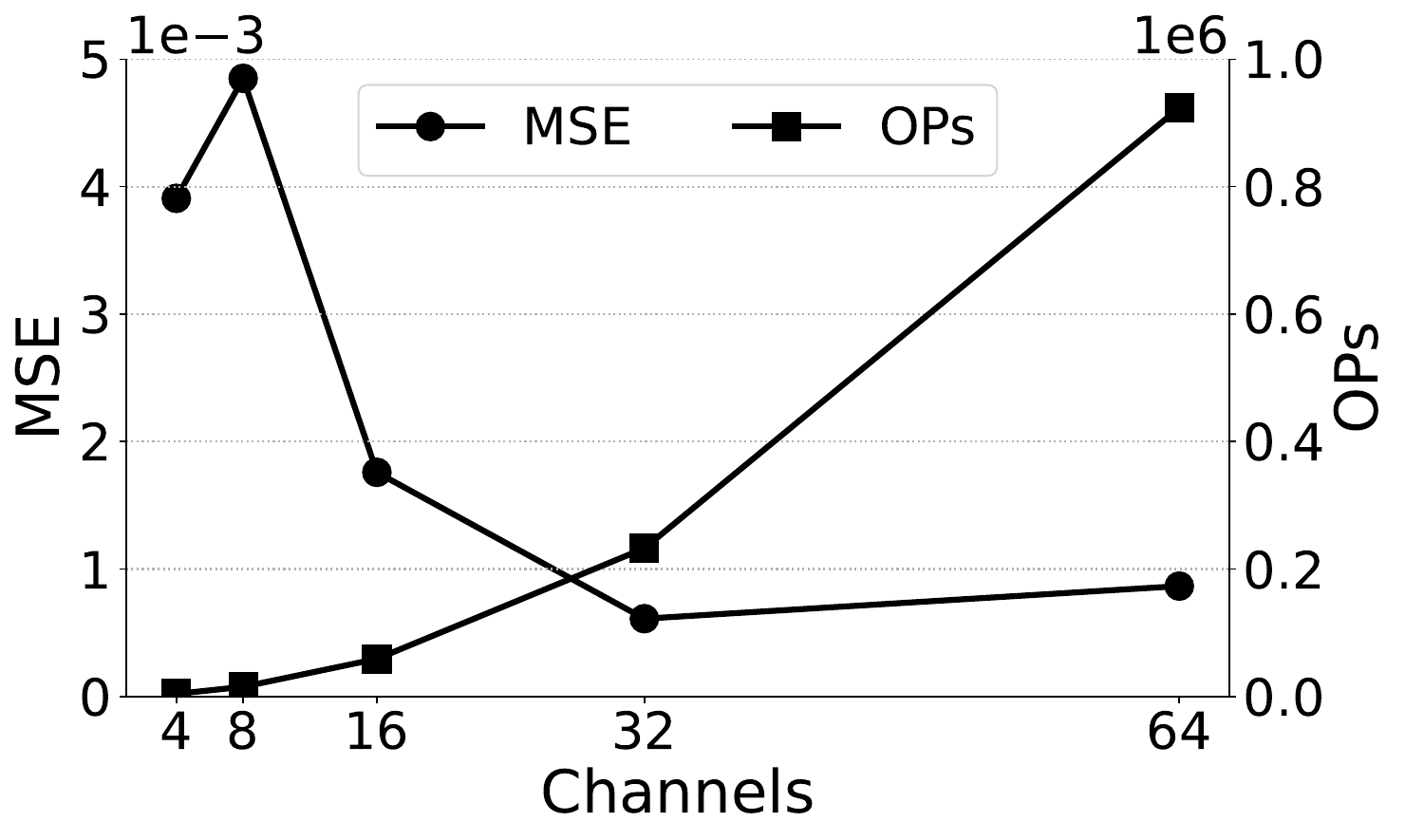}
}
\hfill
\subfloat[Layer Analysis.\label{fig:layers}]{
  \includegraphics[width=0.31\linewidth]{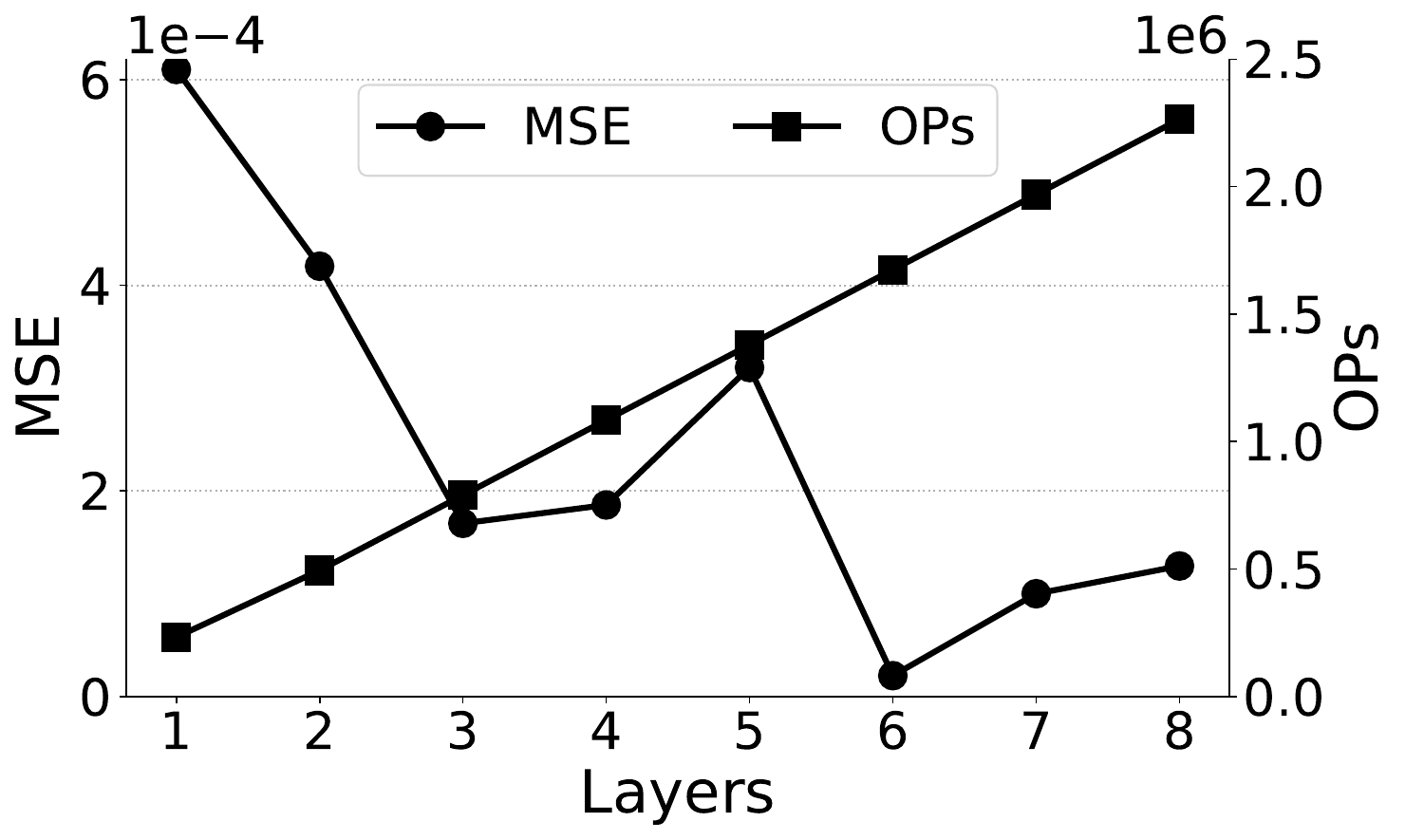}
}
\hfill
\subfloat[Node Analysis.\label{fig:steps}]{
  \includegraphics[width=0.31\linewidth]{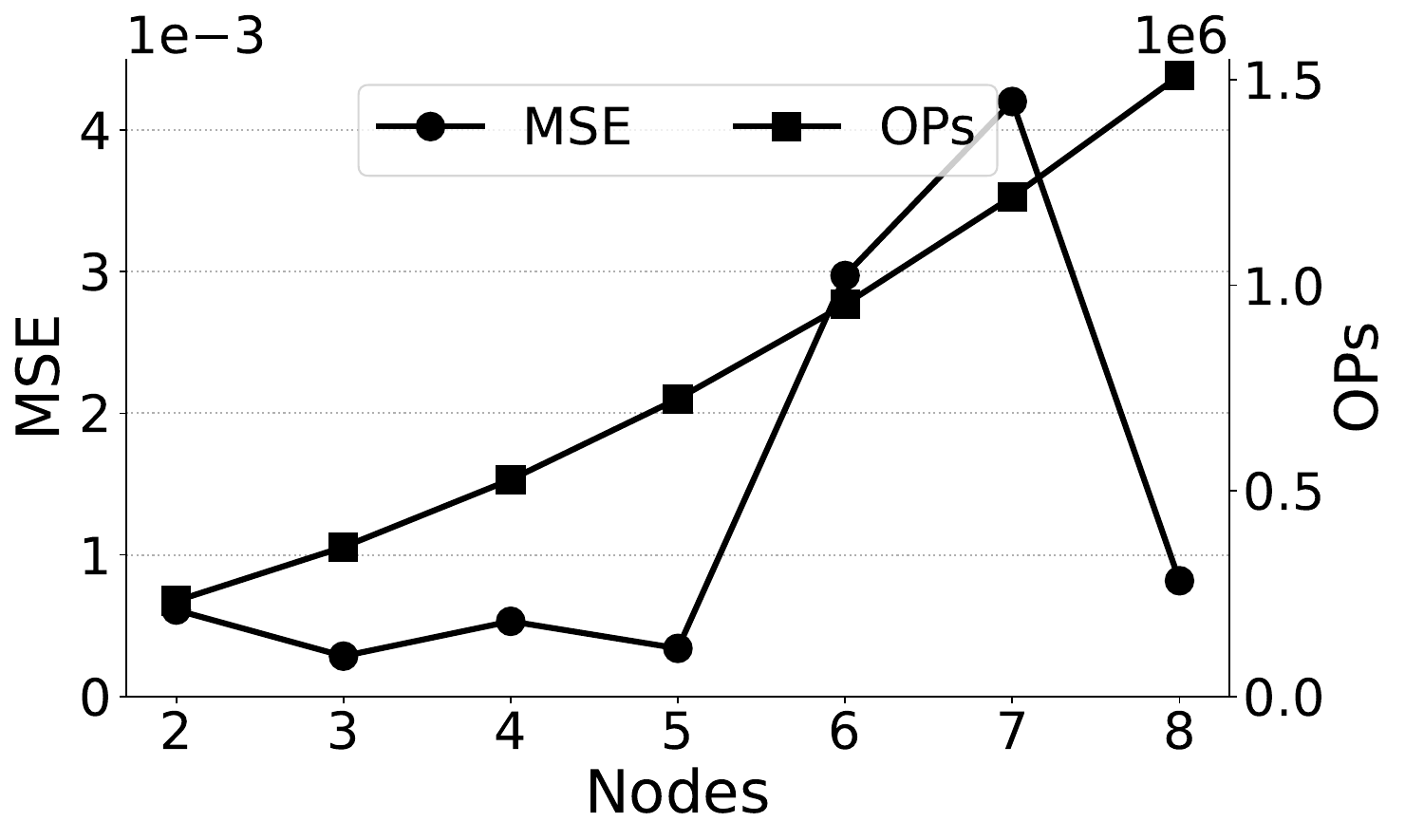}
}
\caption{Comparative Analysis of DARTS Parameters. (a) Channel Analysis: OPs exhibits a quasi-logarithmic relationship with channels (b) Layer Analysis: OPs exhibits a linear relationship with layers (c) Node Analysis: OPs exhibits a linear relationship with internal nodes}
\label{fig:cls_loss_ops} 
\end{figure*}

Since applications may differ in their accuracy and performance requirements, we further explore how the NAS-based approximation can balance approximation accuracy and computational overhead. Using the $sin()$ function as a case study, we derive different approximation models by adjusting the key DARTS configuration parameters $\{C, L, S\}$, where $C$, $L$, and $S$ denote the number of channels, layers, and internal nodes, respectively. As illustrated in Fig.~\ref{fig:cls_loss_ops}, increasing $C$ enhances both accuracy and operations per second (OPS), especially in the range of 16–32 channels, while providing flexible trade-offs between precision and model size (Fig.~\ref{fig:channel}). Increasing network depth $L$ improves approximation accuracy but causes the model size to grow rapidly (Fig.~\ref{fig:layers}). Setting (S = 3) achieves a desirable balance between representational depth and model complexity (Fig.~\ref{fig:steps}).
Overall, the number of operations (OPs) roughly exhibits a logarithmic relationship with the number of channels and a linear relationship with both the number of layers and the number of nodes per cell. Consequently, for memory-constrained hardware, increasing model width (i.e., channels or steps) is generally more favorable than stacking additional layers. Conversely, when higher accuracy is required, deeper architectures yield more substantial improvements.

% \begin{figure*}[htbp]
% \centering
% \begin{subfigure}[b]{0.32\textwidth}
%     \includegraphics[width=\textwidth]{Channels.pdf}
%     \caption{Channel Analysis.}
%     \label{fig:channel}
% \end{subfigure}
% \hfill
% \begin{subfigure}[b]{0.32\textwidth}
%     \includegraphics[width=\textwidth]{Layers.pdf}
%     \caption{Layer Analysis.}
%     \label{fig:layers}
% \end{subfigure}
% \hfill
% \begin{subfigure}[b]{0.32\textwidth}
%     \includegraphics[width=\textwidth]{Nodes.pdf}
%     \caption{Node Analysis.}
%     \label{fig:steps}
% \end{subfigure}
% \caption{Comparative Analysis of DARTS Parameters. (a) Channel Analysis: OPs exhibits a quasi-logarithmic relationship with channels (b) Layer Analysis: OPs exhibits a linear relationship with layers (c) Node Analysis: OPs exhibits a linear relationship with internal nodes}
% \label{fig:{C, L, S} - loss-OPS}
% \end{figure*}

\subsection{Application Approximation Evaluation}

In this experiment, we evaluated the end-to-end performance, energy consumption, and accuracy of both realistic and synthetic applications. Because of the accuracy constraints, we may only have part of the computing kernels in these applications approximated and accelerated. Thus, we set varied proportion of approximal tasks for the four synthetic applications. For the FOC application, it has two approximation models i.e. FOC-NN1 model and FOC-NN2 model to replace the computing kernels of FOC. The detailed models are listed in Table~\ref{tab:realistic_application_models}. Specifically, FOC-NN1 model, approximating the Clarke and Park transformations, achieves a mean squared error (MSE) of 0.002455. FOC-NN2 model, targeting the inverse Park transformation and SVPWM computation, attains an MSE of 0.003697. 
For the Heading application, it utilizes Heading-NN model to replace the computing kernels of Heading. As listed in Table~\ref{tab:realistic_application_models}, Heading-NN model achieves a mean squared error (MSE) of 0.003073. 
Since the synthetic applications do not perform meaningful computation results, we only present their performance and energy consumption in the experiments.

\begin{table}[htbp]
\centering
\caption{Parameters and Performance of the Approximate Kernels in Realistic Applications}
\label{tab:realistic_application_models}

\begin{threeparttable}

\centering
\textbf{(a) Model statistics.}

\resizebox{0.8\linewidth}{!}{
\begin{tabular}{lccc}
\toprule
Model & OPs & MSE & Memory Usage (B) \\
\midrule
FOC-NN1 & 1409024 & 0.002455 & 5344(1.2\%) \\
FOC-NN2 & 1679360 & 0.003697 & 6368(1.4\%) \\
Heading-NN & 1409024 & 0.003073 & 5344(1.2\%) \\
\bottomrule
\end{tabular}
}

\vspace{0.8em}

\textbf{(b) Performance breakdown.}

\resizebox{\linewidth}{!}{
\begin{tabular}{lcccc}
\toprule
Model & Metric & Load Model & Load Input & Inference \\
\midrule
\multirow{2}{*}{FOC-NN1} & Time (\si{\micro\second})   & 4.684 & 0.656 & 11.344 \\
                         & Energy (\si{\micro\joule})  & 0.05  & 0.01  & 0.35 \\
\midrule
\multirow{2}{*}{FOC-NN2} & Time (\si{\micro\second})   & 5.428 & 0.655 & 13.345 \\
                         & Energy (\si{\micro\joule})  & 0.07  & 0.01  & 0.48 \\
\midrule
\multirow{2}{*}{Heading-NN} & Time (\si{\micro\second})   & 5.475 & 1.319 & 13.681 \\
                            & Energy (\si{\micro\joule})  & 0.05  & 0.01  & 0.45 \\
\bottomrule
\end{tabular}
}

\begin{tablenotes}
\footnotesize
\item FOC-NN1 approximates Clarke and Park transformations.
\item FOC-NN2 approximates the inverse Park transformation and SVPWM.
\item Heading-NN approximates the heading estimation.
\end{tablenotes}

\end{threeparttable}
\end{table}

\subsubsection{Performance and Energy Evaluation}
\begin{figure}[htbp]
\centerline{\includegraphics[width=1\linewidth]{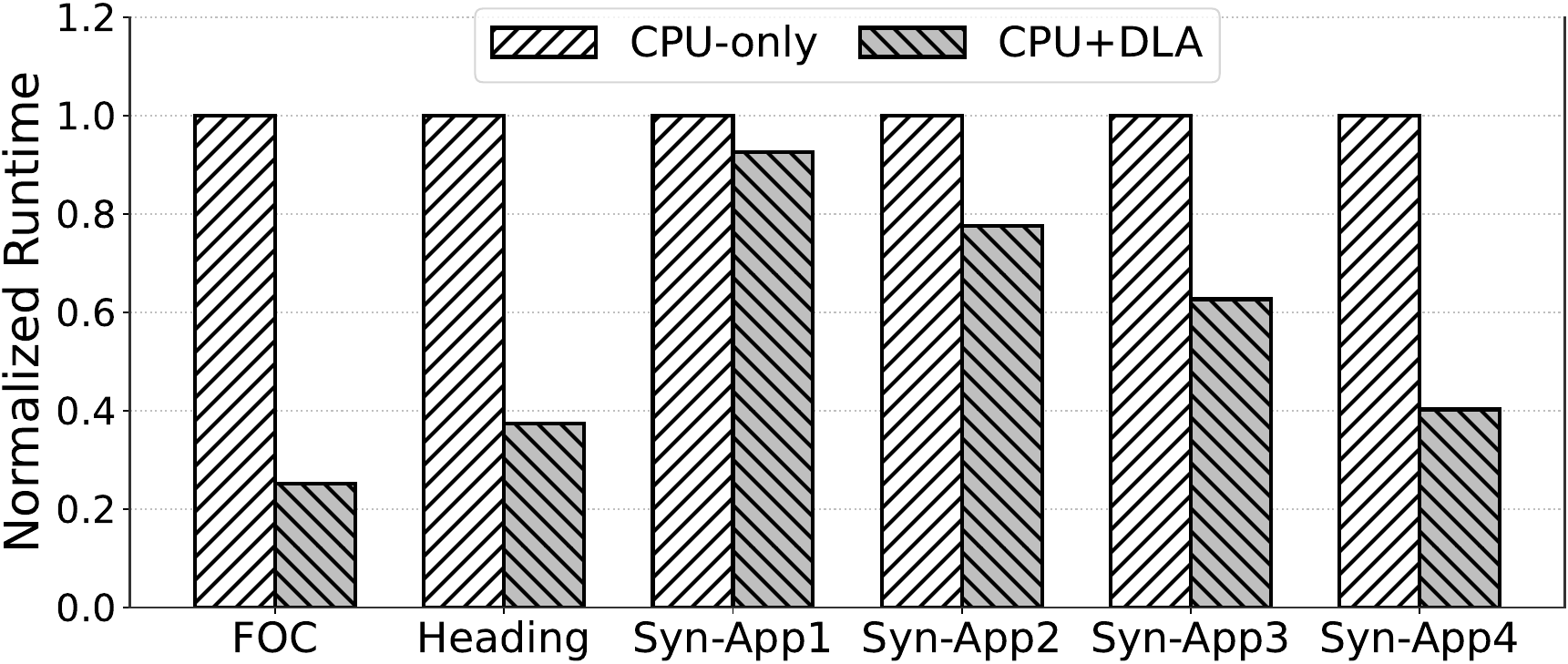}}
\caption{Normalized runtime comparison between CPU-only baselines and CPU-DLA cooperative implementations across representative applications.}
\label{casestudy_runtime}
\end{figure}

\begin{figure}[htbp]
\centerline{\includegraphics[width=1\linewidth]{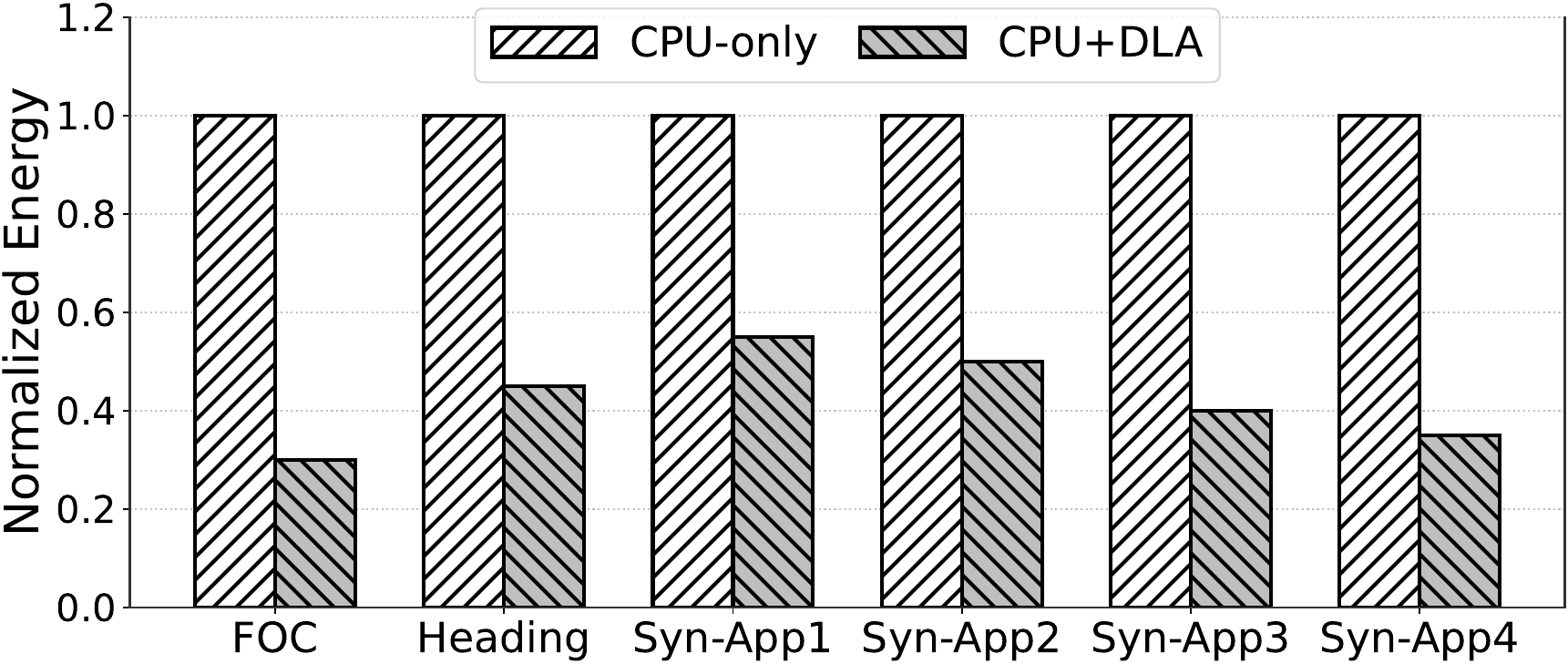}}
\caption{Normalized energy consumption comparison between CPU-only baselines and CPU-DLA cooperative implementations across representative applications.}
\label{casestudy_energy}
\end{figure}

The execution time and energy consumption of the evaluated applications are summarized in Fig.~\ref{casestudy_runtime} and Fig.~\ref{casestudy_energy}, respectively. The experimental results demonstrate that these applications significantly benefit from the approximation of computationally intensive transcendental functions, achieving notable performance speedups and energy savings compared to the baseline CPU-only implementations. As shown in Table~\ref{tab:realistic_application_models}, the approximated kernels executed on the DLA run substantially faster than their default counterparts on embedded CPUs, leading to corresponding reductions in energy consumption. Unlike the kernel-level evaluation in Table~\ref{tab:benchmark}, some of the approximate kernels in these applications are more complex, and the loading overhead becomes less dominant. 

\begin{figure}[htbp]
\centerline{\includegraphics[width=1\linewidth]{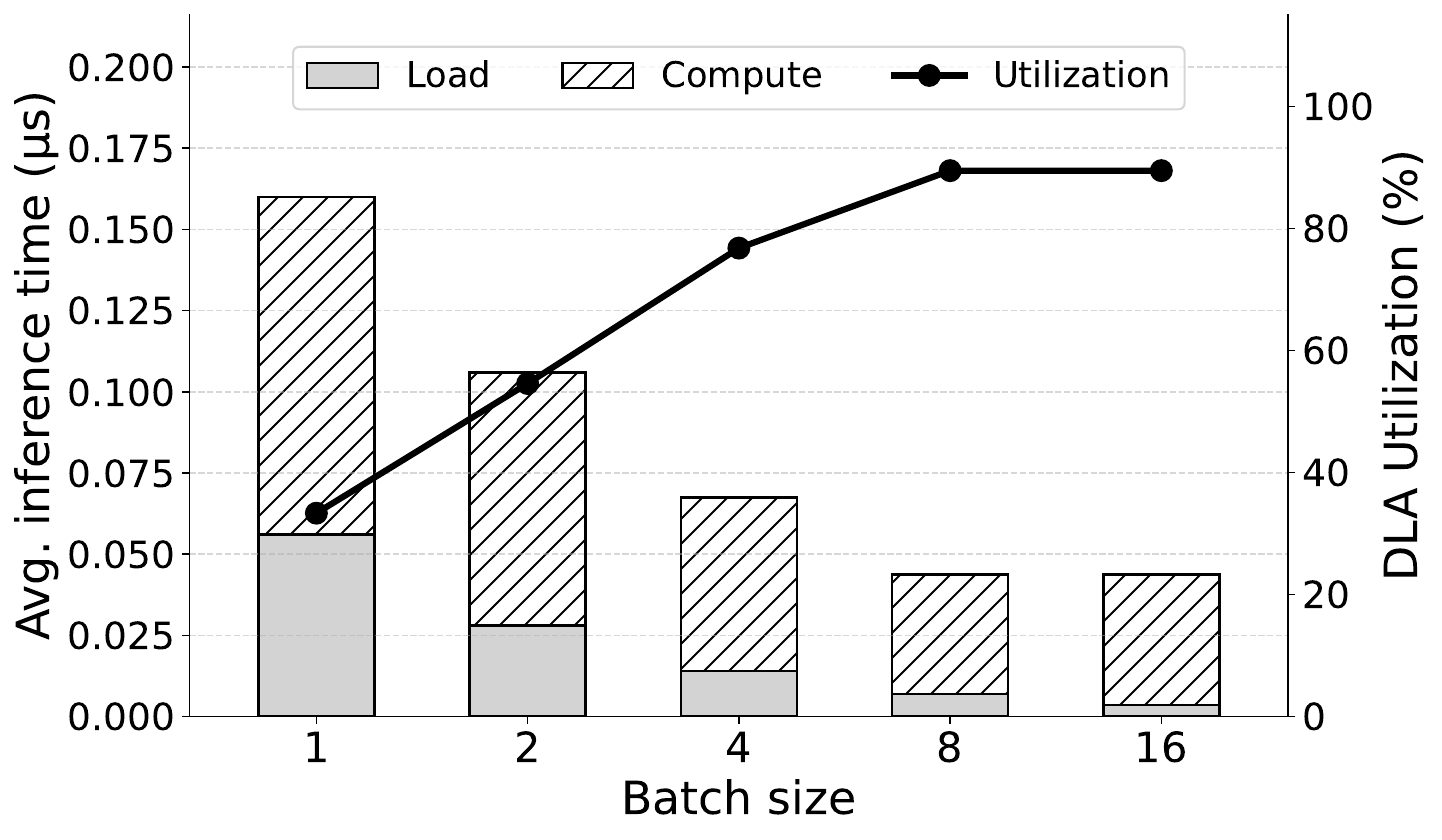}}
\caption{Batch parallelization. As the batch size increases, multiple DLAs operate concurrently and they only need to load the model once. Hence, the average inference latency decreases rapidly. When the average loading overhead becomes a fraction of the inference and the parallel processing units are fully utilized for larger bath size setups, the performance improvement saturates.}
\label{batch_breakdown_time}
\end{figure}

Furthermore, as illustrated in Fig.~\ref{batch_breakdown_time}, the overhead of data loading that essentially transfers both weights and inputs from main memory to DLA can be effectively amortized when multiple inputs are processed in batches. This is because the approximate models are loaded only once and can be reused for different inputs. Thereby, the average loading time per inference is reduced. In addition, when multiple DLAs are provided, the different inputs in the same batch can also be processed in parallel, which can fully exploit the parallel processing capability of the DLAs effectively and further benefit the average inference. For the FOC-NN1 model, the performance improvement closely follows Amdahl’s law, depending on the proportion of tasks that can be approximated and offloaded to the DLA. Greater performance and energy gains can be expected until the DLA becomes saturated. Overall, the proposed AI harvesting strategy reduces application execution time by 72.8\% on average by using the neural network–based kernel approximation and offloading to the DLA.

\subsubsection{Accuracy Evaluation}

As the synthetic applications do not produce meaningful outputs, we evaluate computing accuracy only on the realistic applications. Specifically, for the Field-Oriented Control (FOC) workload—which iteratively adjusts rotor speed via PID control—we implement the algorithm on both CPU-only and CPU+DLA configurations. The CPU-only version employs floating-point arithmetic and serves as the golden reference. In contrast, the CPU+DLA configuration utilizes the proposed AI harvesting strategy to replace computational kernels with neural networks.

\begin{figure}[htbp]
\centerline{\includegraphics[width=1\linewidth]{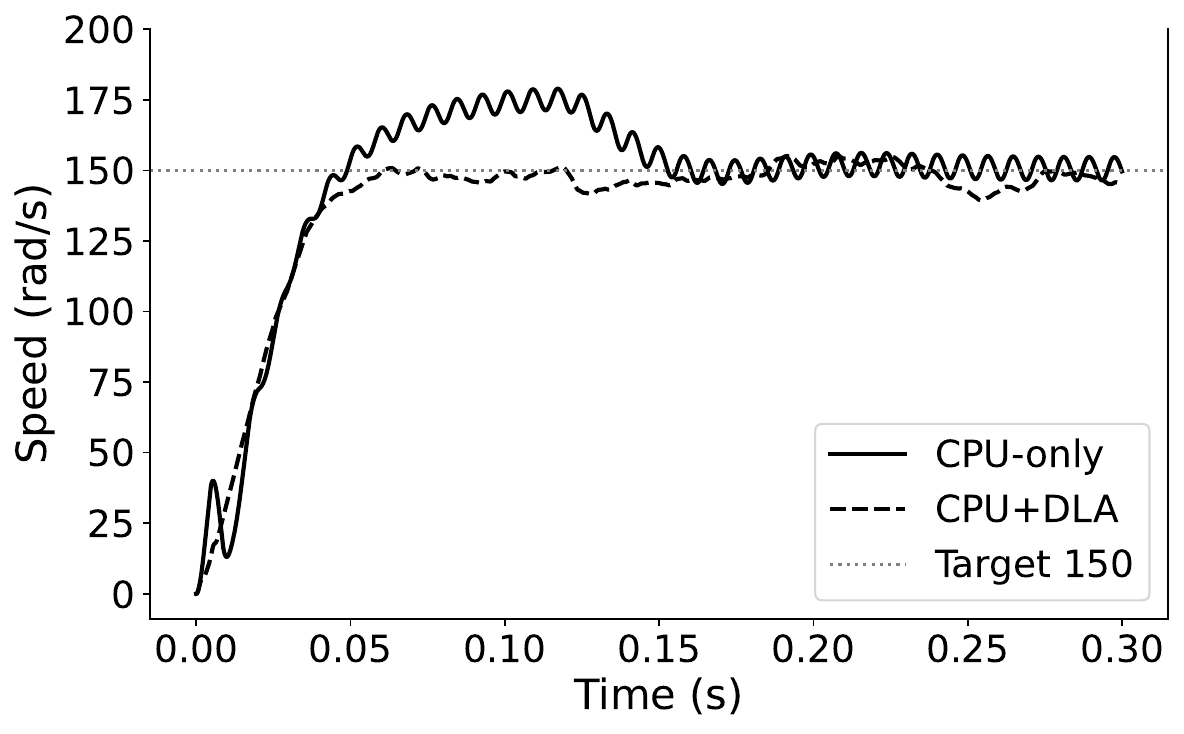}}
\caption{Speed response comparison between the CPU-only and CPU+DLA systems using the same PID parameters. The target speed is set to 150~rad/s.}
\label{foc_sim}
\end{figure}

The experimental results, shown in Fig.~\ref{foc_sim}, illustrate the convergence behavior of the control process. The target speed is set to \(\SI{150}{\radian/\second}\), and an incremental encoder provides the ground-truth rotor speed. Both systems use identical PID gains. The CPU-only system (solid line) exhibits a noticeable fluctuation near \(t \approx \SI{0.01}{\second}\), corresponding to the transient overshoot typical of PID controllers: the proportional term reacts strongly to the large initial error, while the integral term has yet to accumulate. Motor inertia and delays in the inner current-control loop further amplify this transient, producing brief oscillations before convergence.

In contrast, the CPU+DLA system (dashed line) demonstrates a smoother transient response, benefiting from roughly a fourfold increase in control-cycle throughput compared to the CPU-only configuration. As the rotor speed approaches the setpoint, the CPU-only implementation exhibits periodic oscillations during steady-state operation, whereas the CPU+DLA system shows small, irregular ripples caused by approximation errors introduced by the learned model. Nevertheless, the magnitude of these ripples is smaller than the steady oscillations of the CPU-only system. Overall, the CPU+DLA configuration reaches steady state more quickly and achieves lower oscillation amplitude, indicating that the proposed approach improves control stability and responsiveness despite minor approximation-induced perturbations.

%For the EKF-based heading estimation which XXX planar unmanned ground vehicle (UGV) executes a commanded turn (explain the application in general), we also have it implemented on both CPU-only and CPU+DLA configurations, respectively. Similarly, we have floating point arithmetic used for the CPU-only configuration and have neural network models to approximate the computing kernels for the CPU+DLA configuration such that the computing kernels can be accelerated. Meanwhile, we also provide the ground truth which can be obtained from a higher-precision reference instrument. As illustrated in Fig.~\ref{ekf_sim}, even though the CPU+DLA configuration has approximation-induced computing errors in the heading model, it quadruples the loop update rate (from \(T_s \approx \SI{4.134}{\milli\second}\) to \(\SI{1.043}{\milli\second}\)). In this case, during the fast transient, the dominant error arises from temporal effects—zero-order-hold (ZOH) discretization and end-to-end latency—rather than model bias. Approximating the transient slope by \(\dot{\psi} \approx \SI{230}{\radian\per\second}\), the ZOH rms error scales as \(\dot{\psi} T_s / \sqrt{12}\), yielding \(\approx \SI{0.27}{\radian}\) (CPU-only) versus \(\approx \SI{0.07}{\radian}\) (CPU+DLA). Consequently, the higher-rate CPU+DLA system exhibits smaller phase lag and staircase quantization, resulting in lower overall tracking error despite the approximation.

For the EKF-based heading estimation, in which a planar unmanned ground vehicle (UGV) executes a commanded turn, we implement the algorithm on both CPU-only and CPU+DLA configurations. Similar to the previous case, the CPU-only version employs floating-point arithmetic and serves as the reference, while the CPU+DLA configuration leverages neural network models to approximate the computational kernels, thereby accelerating their execution. Ground-truth heading data are obtained from a high-precision reference instrument for accuracy evaluation.

\begin{figure}[htbp]
\centerline{\includegraphics[width=1\linewidth]{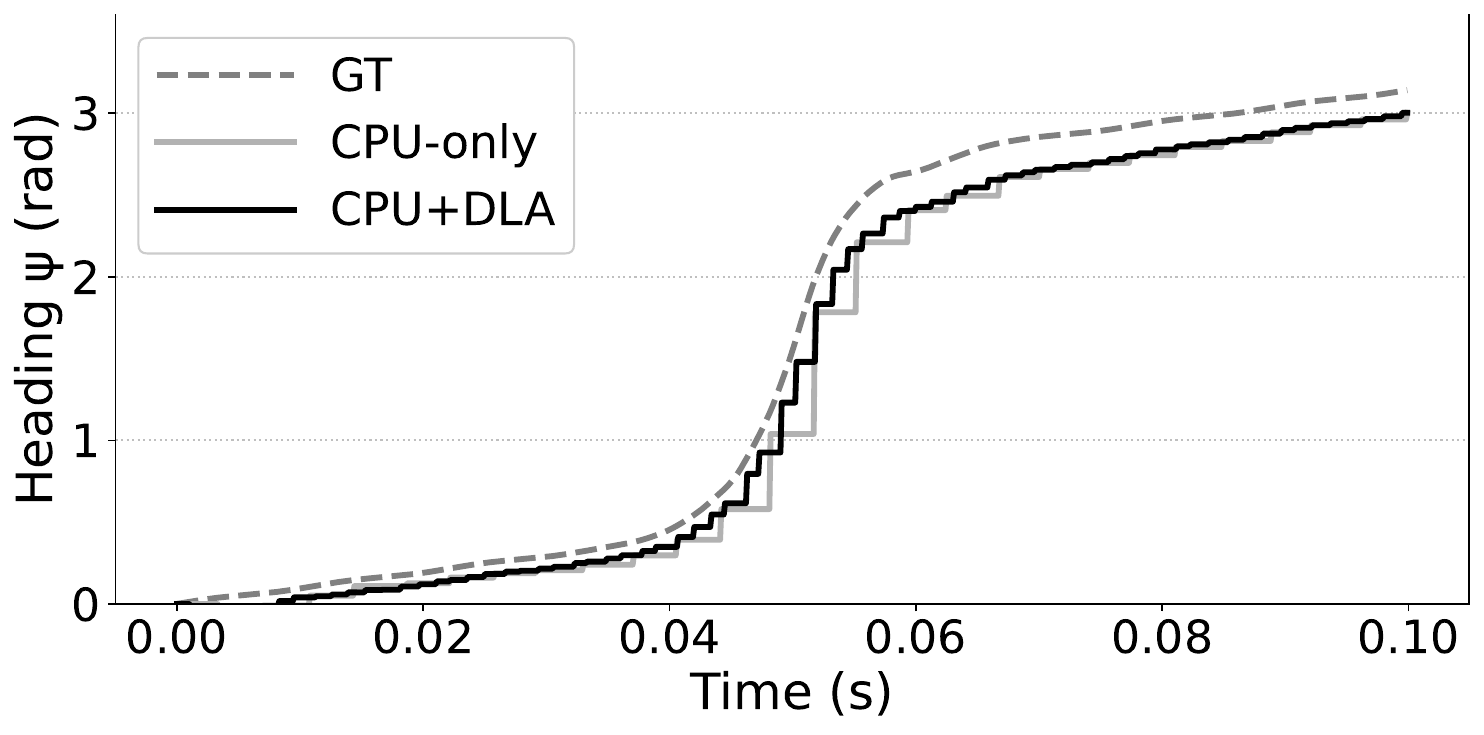}}
\caption{Heading tracking for a planar ground platform. Dashed curve: ground truth (GT). Solid/gray staircases: estimates from CPU-only and CPU+DLA pipelines. Despite a small approximation error in the CPU+DLA model, the $4\times$ higher update rate reduces ZOH/latency-induced temporal error during the rapid turning phase ($\approx$0.045-0.06 s), leading to a lower overall tracking error.}
\label{ekf_sim}
\end{figure}

As illustrated in Fig.~\ref{ekf_sim}, although the CPU+DLA configuration introduces approximation-induced errors in the heading model, it achieves a fourfold increase in loop update rate (from \(T_s \approx \SI{4.134}{\milli\second}\) to \(\SI{1.043}{\milli\second}\)). Under fast-transient conditions, the dominant error arises primarily from temporal effects—zero-order-hold (ZOH) discretization and end-to-end latency—rather than model bias. Approximating the transient slope by \(\dot{\psi} \approx \SI{230}{\radian\per\second}\), the ZOH root-mean-square (rms) error scales as \(\dot{\psi} T_s / \sqrt{12}\), yielding \(\approx \SI{0.27}{\radian}\) for the CPU-only configuration and \(\approx \SI{0.07}{\radian}\) for the CPU+DLA configuration. Consequently, the CPU+DLA system exhibits reduced phase lag and finer temporal resolution, resulting in lower overall tracking error despite the presence of approximation-induced deviations.

In summary, the two realistic applications are found to be more sensitive to computing latency than to approximation-induced errors. The proposed AI harvesting strategy, which leverages neural network approximation to accelerate computation on DLAs, effectively enhances application performance. As a result, even from an accuracy perspective, latency-sensitive applications ultimately benefit from the improved computational efficiency enabled by this approach.

\subsubsection{Scheduling Evaluation}
When applications consist of multiple dependent tasks, executing them sequentially on the CPU and DLA can lead to inefficiencies, as different tasks may favor different computing engines, resulting in load imbalance. Therefore, effective runtime task scheduling is critical for maximizing the utilization of both processing engines. The proposed approximation approach enables flexible task execution on either the CPU or DLA, thereby expanding the scheduling design space and offering greater performance potential. In this experiment, we evaluate both realistic and synthetic applications on the MAX78000 CPU+DLA architecture, comparing the performance of the proposed execution with EFT scheduling against a baseline implementation where CPU-only tasks are executed on the CPU, while CPU/DLA tasks are assigned to randomly selected DLAs following the task’s topological order.

% without scheduling optimization. As shown in Figure.~\ref{schduling}, the proposed runtime scheduling achieves significant performance speedups for applications containing a larger number of tasks, whereas little to no improvement is observed for applications with only two or three tasks—consistent with the limited scheduling flexibility in such cases.
% \begin{figure}[htbp]
% \centerline{\includegraphics [width=1\linewidth]{runtime-tasks_num.pdf}}
% \caption{Runtime comparison between the proposed AI harvesting strategy and baseline implementation in case of streamed processing tasks.}
% \label{schduling}
% \end{figure}

\begin{figure}[htbp]
\centerline{\includegraphics [width=1\linewidth]{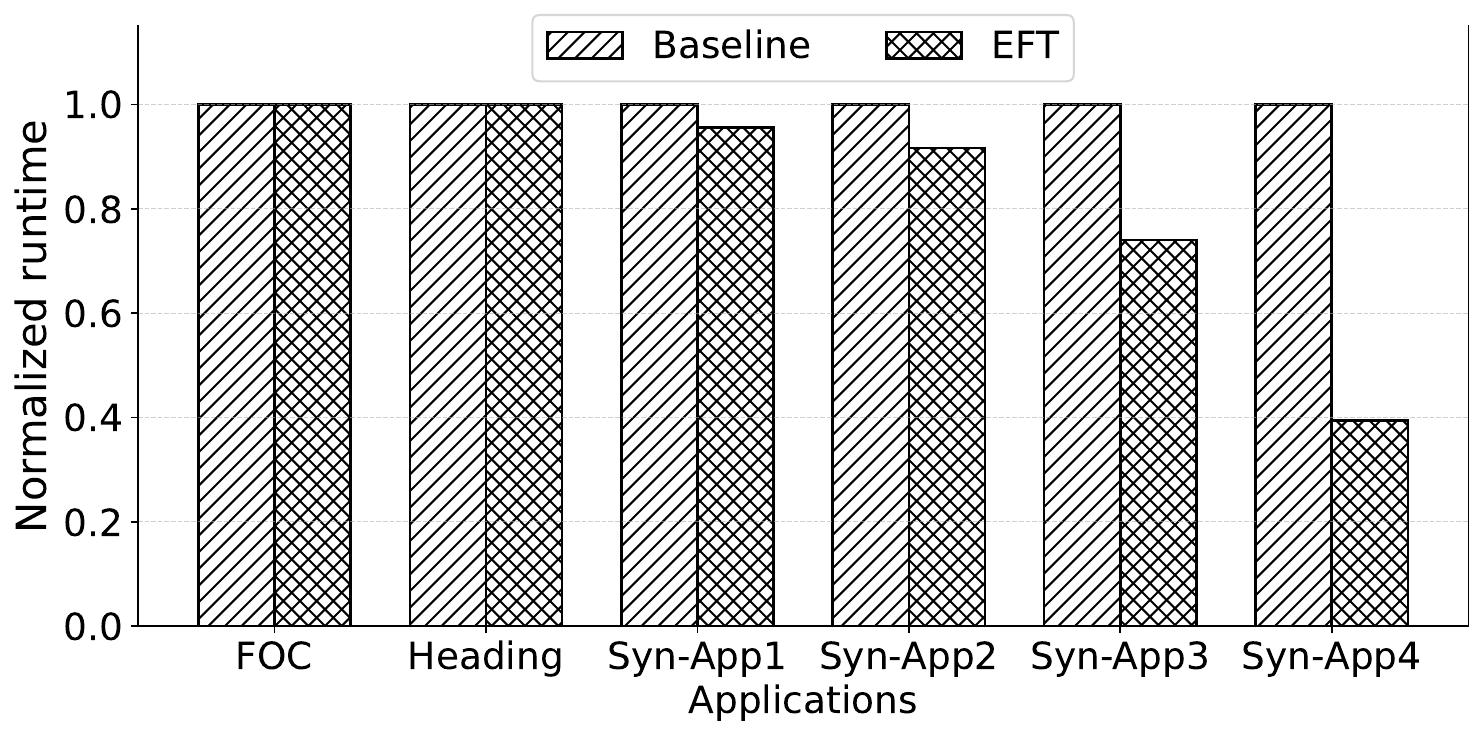}}
\caption{Runtime comparison between the proposed AI harvesting strategy and baseline implementation.}
\label{schduling1}
\end{figure}

As the proposed neural network approximation may not always be applicable or some tasks may be sensitive to approximation errors, the set of tasks that can be approximated varies across different applications. To analyze the impact of the proportion of approximable tasks on system performance, we apply the EFT scheduling strategy to four synthetic applications, and present the results in Fig.~\ref{schduling1}. As shown, the performance advantage of the proposed AI harvesting strategy becomes increasingly significant as the proportion of approximable tasks increases. Specifically, when the proportion is relatively low (e.g., 0.1 in Syn-App1), the EFT method reduces runtime by 4.4\% compared to the baseline. When the proportion increases to 0.5, the runtime reduction reaches 25.9\%, and at 0.8, it further improves to 60.5\%. These results indicate that as the percentage of approximable tasks grows, the AI harvesting strategy can offload more workloads to the DLA and expand the scheduling space, thereby substantially enhancing overall scheduling efficiency and system performance.

\begin{figure}[htbp]
\centerline{\includegraphics [width=1\linewidth]{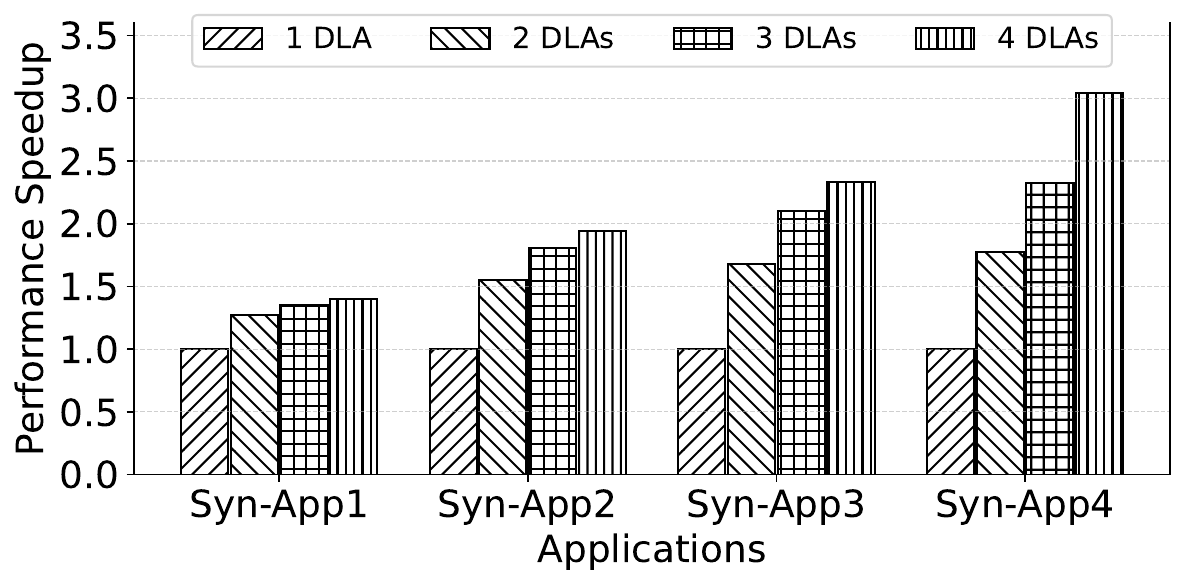}}
\caption{Performance speedup of the proposed AI harvesting strategy implementation in case of different number of AI accelerators.}
\label{schduling2}
\end{figure}

To further examine the impact of accelerator quantity on system performance, we conduct synthetic benchmarks with varying numbers of DLAs. As shown in Fig.~\ref{schduling2}, the overall performance generally improves as the number of DLAs increases, though the rate of improvement differs notably across applications. When the proportion of approximable tasks is low (e.g., Syn-App1 and Syn-App2), adding more DLAs provides limited benefit, and the speedup of Syn-App1 nearly saturates beyond three DLAs due to the dominance of CPU-bound tasks. As the proportion of approximable tasks grows, the benefit of employing multiple DLAs becomes increasingly evident. In particular, when the proportion reaches 0.8, four DLAs achieve over a 3$\times$ speedup compared to the single-DLA configuration, demonstrating the efficiency of the proposed multi-accelerator scheduling strategy under approximation-friendly workloads.
These results confirm that in a multi-accelerator environment, the AI harvesting strategy can effectively distribute tasks across available compute engines, leading to improved utilization of computational resources. By dynamically allocating approximable tasks to accelerators, the system not only alleviates CPU bottlenecks but also enhances overall computational throughput.

\section{Conclusion}

To address the underutilization of AI accelerators in AIoT processors, we propose a novel AI computation harvesting strategy. Specifically, we have developed a comprehensive approximation computing library using NAS, and offload traditional compute-intensive tasks to AI accelerators through neural network approximations, thereby converting idle AI resources into usable computational capacity. To support this strategy, we designed a runtime task scheduler that dynamically allocates tasks to either the CPU or AI accelerators based on task characteristics and system load, without affecting the performance of the primary workload. This approach effectively resolves the underutilization of AI accelerators in both time and space. In complex simulated environments, the framework achieved a performance acceleration of up to 60.5\%, demonstrating the scalability and effectiveness of the proposed strategy. The experimental results highlight the potential application of the framework across a wide range of AIoT systems, enabling more efficient deployment of AI workloads at the edge and significantly improving both performance and energy efficiency.

\newpage

\bibliographystyle{unsrt}
\bibliography{ref}

 

% % \bf{If you include a photo:}\vspace{-33pt}
% % \begin{IEEEbiography}[{\includegraphics[width=1in,height=1.25in,clip,keepaspectratio]{fig1}}]{Michael Shell}
% % Use $\backslash${\tt{begin\{IEEEbiography\}}} and then for the 1st argument use $\backslash${\tt{includegraphics}} to declare and link the author photo.
% % Use the author name as the 3rd argument followed by the biography text.
% % \end{IEEEbiography}


% % \bf{If you will not include a photo:}\vspace{-33pt}
% % \begin{IEEEbiographynophoto}{John Doe}
% % Use $\backslash${\tt{begin\{IEEEbiographynophoto\}}} and the author name as the argument followed by the biography text.
% % \end{IEEEbiographynophoto}

% \vfill

\end{document}